\documentstyle[amssymb,preprint,aps]{revtex}

\begin{document}
\title{{\LARGE Hydrodynamical formulation of quantum mechanics, K\"{a}hler
structure, and Fisher information}}
\author{Marcel Reginatto}
\address{Environmental Measurements Laboratory, U. S. Department of Energy\\
201 Varick St., 5th floor, New York, New York 10014-4811, USA}
\date{\today}
\maketitle

\begin{abstract}
The Schr\"{o}dinger equation can be derived using the minimum Fisher
information principle. \ I discuss why such an approach should work, and
also show that the K\"{a}hler and Hilbert space structures of quantum
mechanics result from combining the symplectic structure of the
hydrodynamical model with the Fisher information metric.


\end{abstract}

\section{Introduction}

In a previous paper \cite{Reginatto98}, it was shown that the hydrodynamical
formulation of the Schr\"{o}dinger equation can be derived using an
information-theoretical approach that is based on the principle of minimum
Fisher information. \ A derivation along similar lines is also possible for
other non-relativistic quantum mechanical equations, such as the Pauli
equation \cite{Reginatto99} and the equation for the quantum rotator \cite
{Reginatto00}. \ The purpose of this paper is two-fold: to examine why such
an information-theoretical approach should work, and to show that the
K\"{a}hler and Hilbert space structures of quantum mechanics result from
combining the symplectic structure of the hydrodynamical model with the
Fisher information metric of information theory. \ The complex
transformation of the hydrodynamical variables that puts this K\"{a}hler
metric in its canonical form is the one that leads to the usual
Schr\"{o}dinger representation.

Frieden \cite{Frieden88}\ was the first one to point out a connection
between the principle of minimum Fisher information and the Schr\"{o}dinger
equation. \ Frieden and coworkers later developed and extended this work in
a series of papers which made use of a new principle called the extreme
physical information (EPI) principle. \ In this paper I will not discuss the
EPI principle, which differs from the principle of minimum Fisher
information in many ways (for a review of the EPI approach, see the book by
Frieden \cite{Frieden99}), but will concentrate instead on the
information-theoretical approach used in \cite{Reginatto98}. \ In this
approach, the emphasis is on using the principle of minimum Fisher
information to complement a physical picture derived from a hydrodynamical
model. \ Applying the principle under the assumption that one can describe
the motion of particles in terms of a hydrodynamical model leads directly to
Madelung's hydrodynamical formulation of quantum mechanics \cite{Madelung26}.

\section{Cross-entropy and Fisher information}

Let $P(y^{i})$ be a probability density which is a function of $n$
continuous coordinates $y^{i}$, and let $P(y^{i}+\Delta y^{i})$ be the
density that results from a small change in the $y^{i}$. Expand the $%
P(y^{i}+\Delta y^{i})$ in a Taylor series, and calculate the cross-entropy $%
J $ up to the first non-vanishing term, 
\begin{eqnarray}
\left. J(P(y^{i}+\Delta y^{i}):P(y^{i}))\right. &=&\int P(y^{i}+\Delta
y^{i})\ln \frac{P(y^{i}+\Delta y^{i})}{P(y^{i})}d^{n}y  \label{Ijk} \\
&\simeq &\left[ \frac{1}{2}\int \frac{1}{P(y^{i})}\frac{\partial P(y^{i})}{%
\partial y^{j}}\frac{\partial P(y^{i})}{\partial y^{k}}d^{n}y\right] \Delta
y^{j}\Delta y^{k}  \nonumber \\
&=&I_{jk}\Delta y^{j}\Delta y^{k}  \nonumber
\end{eqnarray}
The $I_{jk}$ are the elements of the Fisher information matrix. \ This is
not the most general expression for the Fisher information matrix, but the
particular case that is of interest here. \ The general expression is of the
form \cite{Kullback59} 
\begin{equation}
I_{jk}(\theta ^{i})=\frac{1}{2}\int \frac{1}{P(x^{i}|\theta ^{i})}\frac{%
\partial P(x^{i}|\theta ^{i})}{\partial \theta ^{j}}\frac{\partial
P(x^{i}|\theta ^{i})}{\partial \theta ^{k}}d^{n}x  \label{FI}
\end{equation}
where $P(x^{i}|\theta ^{i})$ is a probability density that depends on a set
of $n$\ parameters $\theta ^{i}$ in addition to the $n$ coordinates $x^{i}$.
The expression for the $I_{jk}$\ that appears in equation (\ref{Ijk}) can be
derived from the general formula if 
\[
P(x^{i}|\theta ^{i})=P(x^{i}+\theta ^{i}). 
\]
To see this, introduce a new set of parameters $y^{i}=x^{i}+\theta ^{i}$. \
Then 
\[
I_{jk}(\theta ^{i})\rightarrow \frac{1}{2}\int \frac{1}{P(y^{i})}\frac{%
\partial P(y^{i})}{\partial y^{j}}\frac{\partial P(y^{i})}{\partial y^{k}}%
d^{n}y=I_{jk} 
\]
since $d^{n}x\rightarrow d^{n}y$ as the integration over the $x^{i}$
coordinates is for fixed values of $\theta ^{i}$.

If $P$\ is defined over an $n$-dimensional manifold $M$ with (positive)
inverse metric $g^{ik}$, there is a natural definition of the amount of
information $I$ associated with $P$, which is obtained by contracting $%
g^{ik} $ with the elements of the Fisher information matrix, 
\begin{equation}
I=g^{ik}I_{ik}=g^{ik}\frac{1}{2}\int \frac{1}{P}\frac{\partial P}{\partial
y^{i}}\frac{\partial P}{\partial y^{k}}d^{n}y.  \label{FIndim}
\end{equation}
The case of interest here is the one where $M$\ is the $n+1$ dimensional
extended configuration space $QT$ (with coordinates $\{t,x^{1},...,x^{n}\}$)
of a non-relativistic particle of mass $m$. \ Then, the inverse metric is
the one used to define the kinematical line element in configuration space,
which is of the form $g^{ik}=diag(0,1/m,...,1/m)$. \ Sometimes it will be
convenient to use quantities defined over the configuration space $Q$ (with
coordinates $\{x^{1},...,x^{n}\}$) rather than $QT$, and I will do so if it
simplifies the notation.

\section{Derivation of the Schr\"{o}dinger equation}

In the Hamilton-Jacobi formulation of classical mechanics, the equation of
motion takes the form 
\begin{equation}
\frac{\partial S}{\partial t}+\frac{1}{2}g^{\mu \nu }\frac{\partial S}{%
\partial x^{\mu }}\frac{\partial S}{\partial x^{\nu }}+V=0  \label{CHJ}
\end{equation}
where $g^{\mu \nu }=diag(1/m,...,1/m)$\ \cite{Synge}\ is the inverse metric
used to define the kinematical line element in the configuration space $Q$\
parametrized by coordinates $\{x^{\mu }\}$. \ The velocity field $u^{\mu }$
is derived from $S$ according to 
\begin{equation}
u^{\mu }=g^{\mu \nu }\frac{\partial S}{\partial x^{\nu }}.  \label{VEL}
\end{equation}
When the exact coordinates that describe the state of the classical system
are unknown, one usually describes the system by means of a probability
density $P(t,x^{\mu })$. \ The probability density must satisfy the
following two conditions: it must be normalized, 
\[
\int Pd^{n}x=1, 
\]
and it must satisfy a continuity equation, 
\begin{equation}
\frac{\partial }{\partial t}P+\frac{\partial }{\partial x^{\mu }}\left(
Pg^{\mu \nu }\frac{\partial S}{\partial x^{\nu }}\right) =0.  \label{CONT}
\end{equation}
Equations (\ref{CHJ}) and (\ref{CONT}), together with (\ref{VEL}),
completely determine the motion of the classical ensemble. \ Equations (\ref
{CHJ}) and (\ref{CONT}) can be derived from the Lagrangian 
\begin{equation}
L_{CL}=\int P\left\{ \frac{\partial S}{\partial t}+\frac{1}{2}g^{\mu \nu }%
\frac{\partial S}{\partial x^{\mu }}\frac{\partial S}{\partial x^{\nu }}%
+V\right\} dtd^{n}x  \label{CL}
\end{equation}
by fixed end-point variation ($\delta P=\delta S=0$ at the boundaries) with
respect to $S$ and $P$.

Quantization of the classical ensemble is achieved by adding to the
classical Lagrangian (\ref{CL})\ a term proportional to the information $I$
defined by equation (\ref{FIndim}) \cite{Reginatto98}. \ This leads to the
Lagrangian for the Schr\"{o}dinger equation, 
\begin{eqnarray}
L_{QM} &=&L_{CL}+\lambda I  \label{QL} \\
&=&\int P\left\{ \frac{\partial S}{\partial t}+\frac{1}{2}g^{\mu \nu }\left[ 
\frac{\partial S}{\partial x^{\mu }}\frac{\partial S}{\partial x^{\nu }}%
+\lambda \frac{1}{P^{2}}\frac{\partial P}{\partial x^{\mu }}\frac{\partial P%
}{\partial x^{\nu }}\right] +V\right\} dtd^{n}x.  \nonumber
\end{eqnarray}
Fixed end-point variation with respect to $S$ leads again to (\ref{CONT}),
while fixed end-point variation with respect to $P$ leads to 
\begin{equation}
\frac{\partial S}{\partial t}+\frac{1}{2}g^{\mu \nu }\left[ \frac{\partial S%
}{\partial x^{\mu }}\frac{\partial S}{\partial x^{\nu }}+\lambda \left( 
\frac{1}{P^{2}}\frac{\partial P}{\partial x^{\mu }}\frac{\partial P}{%
\partial x^{\nu }}-\frac{2}{P}\frac{\partial ^{2}P}{\partial x^{\mu
}\partial x^{\nu }}\right) \right] +V=0  \label{QHJEQ}
\end{equation}
Equations (\ref{CONT}) and (\ref{QHJEQ}) are identical to the
Schr\"{o}dinger equation provided the wave function $\psi (t,x^{\mu })$ is
written in terms of $S$ and $P$ by 
\[
\psi =\sqrt{P}\exp (iS/\hbar ) 
\]
and the parameter $\lambda $ is set equal to 
\[
\lambda =\left( \frac{\hbar }{2}\right) ^{2}. 
\]
Note that the classical limit of the Schr\"{o}dinger theory is not the
Hamilton-Jacobi equation for a classical particle, but the equations (\ref
{CHJ}) and (\ref{CONT}) which describe a classical ensemble.

It can be shown (see Appendix) that the Fisher information $I$\ increases
when $P$ is varied while $S$ is kept fixed. \ Therefore, the solution
derived here is the one that minimizes the Fisher information for a given $S$%
.

The approach followed here is of interest in that it provides a way of
distinguishing between physical and information-theoretical assumptions (for
a very clear account of the importance of making this type of distinction in
quantum mechanics see the paper by Jaynes \cite{Jaynes89}). In general
terms, the information-theoretical content of the theory lies in the
prescription to minimize the Fisher information associated with the
probability distribution that describes the position of particles, while the
physical content of the theory is contained in the assumption that one can
describe the motion of particles in terms of a hydrodynamical model.

\section{On the use of the minimum Fisher information principle in quantum
mechanics}

The cross-entropy $J$, 
\[
J(Q:P)=\int Q(y^{i})\ln \left( \frac{Q(y^{i})}{P(y^{i})}\right) d^{n}y. 
\]
where $P$, $Q$ are two probability densities, plays a central role in
information theory and in the theory of inference. \ It has properties that
are desirable for an information measure \cite{Kullback59},\ and it can be
argued that it measures the amount of information needed to change a prior
probability density $P$ into the posterior $Q$ \cite{Hobson72}. \
Maximization of the relative entropy (which is defined as the {\it negative}
of the cross-entropy\footnote{%
A note on terminology: due to the connection between relative entropy and
cross-entropy, the {\it maximum entropy principle} is also known as the {\it %
minimum cross-entropy principle}, which can lead to some confusion. \ The
cross-entropy (or its negative) may found in the literature under various
names: Kullback-Leibler information, directed divergence, discrimination
information, Renyi's information gain, expected weight of evidence, entropy,
entropy distance.}) is the basis of the maximum entropy principle, a method
for inductive inference that leads to a posterior distribution given a prior
distribution and new information in the form of expected values. \ The
maximum entropy principle asserts that of all the probability densities that
are consistent with the new information, the one which has the maximum
relative entropy is the one that provides the most unbiased representation
of our knowledge of the state of the system. \ There are several approaches
that lead to the maximum entropy principle. \ In the original derivation by
Jaynes \cite{Jaynes57}, the use of the maximum entropy principle was
justified on the basis of the relative entropy's unique properties as an
uncertainty measure. \ An independent justification based on consistency
arguments was later given by Shore and Johnson \cite{ShoJohn80}. \ Jaynes
had already remarked that inferences made using any other information
measure than the entropy may lead to contradictions. Shore and Johnson
considered the consequences of requiring that methods of inference be
self-consistent. \ They introduced a set of axioms that were all based on
one fundamental principle: if a problem can be solved in more than one way,
the results should be consistent. \ They showed that given information in
the form of a set of constraints on expected values, there is only one
distribution satisfying the set of constraints which can be chosen using a
procedure that satisfies their axioms, and this unique distribution can be
obtained by maximizing the relative entropy. \ Therefore, they concluded
that if a method of inference is based on a variational principle,
maximizing any function but the relative entropy will lead to
inconsistencies unless that function and the relative entropy have identical
maxima (any monotonic function of the relative entropy will work, for
example).

It is tempting to argue by analogy that the minimum Fisher information
derivation of the Schr\"{o}dinger equation is in essence nothing but a
variation on maximum entropy, one in which maximization of relative entropy
is simply replaced by minimization of the Fisher information (some
similarities and differences of the two approaches were discussed briefly in 
\cite{Reginatto98}). \ But if we take into consideration the unique
properties that make cross-entropy {\it the} fundamental measure of
information together with the result of Shore and Johnson, it becomes
difficult to justify a principle of inference based on information theory
that would operate along the same lines as maximum entropy but using the
principle of minimum Fisher information instead. \ To understand the use of
the minimum Fisher information principle in the context of quantum
mechanics, it is crucial to take into consideration that here one is
selecting those probability distributions $P(y^{i})$ for which a
perturbation that leads to $P(y^{i}+\Delta y^{i})$ will result in the
smallest increase of the cross-entropy for a given $S(y^{i})$. \ In other
words, the method of choosing $P(y^{i})$ is based on the idea that a
solution should be stable under perturbations in the very precise sense that
the amount of additional information needed to describe the change in the
solution should be as small as possible. \ We have then a new principle:
choose the probability densities that describe the quantum system on the
basis of the stability of those solutions, {\it where the measure of the
stability is given by the amount of information needed to change }$P(y^{i})$%
{\it \ into }$P(y^{i}+\Delta y^{i})$. \ Why should restricting the choice of 
$\{P,S\}$ to those that are stable in this sense lead to the excellent
predictions of quantum mechanics? \ Such an approach should work for
physical systems that can be represented by models in which the probability
density $P$ describes the equilibrium density of an underlying stochastic
process (see for example the derivation of the diffusion equation using the
minimum Fisher information principle in \cite{RegiLeng99}). \ Such models of
quantum mechanics do exist: a formulation along these lines was first
proposed by Bohm and Vigier \cite{BomVig54}, and later a different but
related formulation was given by Nelson \cite{Nelson66}(for a review of the
stochastic formulation of the quantum theory that compares these two
approaches, see \cite{BohmHi89}). \ Whether the additional assumptions
needed to build these particular models are sound, and whether they provide
a correct description of quantum mechanics will depend of course on the
experimental predictions that they make. \ The minimum Fisher information
approach can be of no help here, since it is only concerned with making
inferences about probability distributions and operates therefore at the
epistemological level.

\section{K\"{a}hler and Hilbert space structures of quantum mechanics}

I now want to examine the assumptions that are needed to construct the
K\"{a}hler and Hilbert space structures of quantum mechanics. \ My aim is
not to give a mathematically rigorous derivation of these results, but to
give arguments that justify introducing the K\"{a}hler space structure on
the basis of mathematical structures that arise naturally in the
hydrodynamical model and in information theory. \ In particular, I want to
show that the K\"{a}hler structure of quantum mechanics results from
combining the symplectic structure of the hydrodynamical model with the
Fisher information metric of information theory. \ The complex
transformation of the hydrodynamical variables that puts this K\"{a}hler
metric in its canonical form is the one that leads to the usual
Schr\"{o}dinger representation. \ Good descriptions of the geometrical
formulation of quantum mechanics covering the case of infinite-dimensional
K\"{a}hler manifolds are available in the literature; see for example
Cirelli et. al.\cite{CiManPi90}, Ashtekar and Schilling \cite{AshSchill99}\
and Brody and Hughston \cite{BrodyHugh97}. \ The approach of Brody and
Hughston is of special interest in that they make explicit use of the Fisher
information metric, although without making reference to the hydrodynamical
formulation.

I first look at the symplectic structure of the hydrodynamical formulation.
\ Introduce as basic variables the hydrodynamical fields $\{P,S\}$. \ The
symplectic structure is given by the two form 
\begin{eqnarray*}
\omega (\delta P(x^{\mu }),\delta S(x^{\mu });\delta ^{\prime }P(x^{\mu
}),\delta ^{\prime }S(x^{\mu })) &=&\int \left\{ \left( \delta P(x^{\mu
}),\delta S(x^{\mu })\right) \left( 
\begin{array}{cc}
0 & 1 \\ 
-1 & 0
\end{array}
\right) \left( 
\begin{array}{c}
\delta ^{\prime }P(x^{\mu }) \\ 
\delta ^{\prime }S(x^{\mu })
\end{array}
\right) \right\} d^{n}x \\
&=&\int \left\{ \left( \delta P(x^{\mu }),\delta S(x^{\mu })\right) \cdot
\Omega \cdot \left( 
\begin{array}{c}
\delta ^{\prime }P(x^{\mu }) \\ 
\delta ^{\prime }S(x^{\mu })
\end{array}
\right) \right\} d^{n}x
\end{eqnarray*}
where $\delta $\ and $\delta ^{\prime }$\ are two generic systems of
increments for the phase-space variables. \ The Poisson brackets for two
functions ${\cal F}^{1}(P,S),$\ ${\cal F}^{2}(P,S)$\ take the form 
\[
\left\{ {\cal F}^{1}(P,S),{\cal F}^{2}(P,S)\right\} =\int \left\{ \left[
\delta {\cal F}^{1}/\delta P\right] \left[ \delta {\cal F}^{2}/\delta S%
\right] -\left[ \delta {\cal F}^{1}/\delta S\right] \left[ \delta {\cal F}%
^{2}/\delta P\right] \right\} d^{n}x. 
\]
The equations of motion (\ref{CONT}), (\ref{QHJEQ}) can be written as 
\begin{eqnarray*}
\frac{\partial P}{\partial t} &=&\left\{ P,{\cal H}\right\} =\frac{\delta 
{\cal H}}{\delta S} \\
\frac{\partial S}{\partial t} &=&\left\{ S,{\cal H}\right\} =-\frac{\delta 
{\cal H}}{\delta P}
\end{eqnarray*}
with the Hamiltonian ${\cal H}$ given by 
\[
{\cal H}=\int P\left\{ \frac{1}{2}g^{\mu \nu }\left[ \frac{\partial S}{%
\partial x^{\mu }}\frac{\partial S}{\partial x^{\nu }}+\left( \frac{\hbar }{2%
}\right) ^{2}\frac{1}{P^{2}}\frac{\partial P}{\partial x^{\mu }}\frac{%
\partial P}{\partial x^{\nu }}\right] +V\right\} d^{n}x. 
\]
${\cal H}$ acts as the generator of time translations.

To introduce the Fisher information metric, let $\theta ^{\mu }$ be a set of
real continuous parameters, and consider the parametric family of positive
distributions defined by 
\[
P(x^{\mu }|\theta ^{\mu })=P(x^{\mu }+\theta ^{\mu }) 
\]
where the probability densities $P$ are solutions of the Schr\"{o}dinger
equation (at time $t=0$). \ Then there is a natural metric over the space of
parameters $\theta ^{\mu }$ given by the Fisher information matrix \cite
{Rao45}, and it leads to a concept of distance defined by 
\begin{equation}
ds^{2}(\theta ^{\mu })=\frac{1}{2}\left[ \int \frac{1}{P(x^{\mu }|\theta
^{\mu })}\frac{\partial P(x^{\mu }|\theta ^{\mu })}{\partial \theta ^{\rho }}%
\frac{\partial P(x^{\mu }|\theta ^{\mu })}{\partial \theta ^{\sigma }}d^{n}x%
\right] \delta \theta ^{\rho }\delta \theta ^{\sigma }  \label{ds2theta}
\end{equation}
Using 
\[
\delta P=\frac{\partial P}{\partial \theta ^{\mu }}\delta \theta ^{\mu } 
\]
one can write equation (\ref{ds2theta}) as 
\begin{equation}
ds^{2}(\theta ^{\mu })=\frac{1}{2}\left[ \int \frac{1}{P(x^{\mu }|\theta
^{\mu })}\delta P(x^{\mu }|\theta ^{\mu })\delta P(x^{\mu }|\theta ^{\mu
})d^{n}x\right]  \label{ds2P}
\end{equation}

We use equation (\ref{ds2P}) to introduce a metric over the space of
solutions of the Schr\"{o}dinger equation (i.e., $P(x^{\mu }|\theta ^{\mu })$
with $\theta ^{\mu }=0$) by setting 
\begin{eqnarray*}
ds^{2}(\delta P,\delta ^{\prime }P) &=&\frac{1}{2}\left[ \int \frac{1}{%
P(x^{\mu })}\delta P(x^{\mu })\delta ^{\prime }P(x^{\mu })d^{3}x\right] \\
&=&\int g^{(P)}\delta P(x^{\mu })\delta ^{\prime }P(x^{\mu })d^{3}x
\end{eqnarray*}
where 
\[
P(x^{\mu })=P(x^{\mu }|\theta ^{\mu }=0), 
\]
\[
\delta P(x^{\mu })=\delta P(x^{\mu }|\theta ^{\mu })|_{\theta ^{\mu }=0} 
\]
\[
g^{(P)}=\frac{1}{2P(x^{\mu })} 
\]

I now want to extend the metric $g^{(P)}$ over the probability densities to
a metric $g_{ab}$ over the whole space $\{P,S\}$ of solutions of the
Schr\"{o}dinger equation, in such a way that the metric structure is
compatible with the symplectic structure. \ To do this, introduce a complex
structure $J_{\ b}^{a}$ and impose the following conditions, 
\begin{equation}
\Omega _{ab}=g_{ac}J_{\ b}^{c}  \label{c1}
\end{equation}
\begin{equation}
J_{\ c}^{a}g_{ab}J_{\ d}^{b}=g_{cd}  \label{c2}
\end{equation}
\begin{equation}
J_{\ b}^{a}J_{\ c}^{b}=-\delta _{\ \ c}^{a}  \label{c3}
\end{equation}
A set of $\{\Omega _{ab},g_{ab},J_{\ b}^{a}\}$\ that satisfy equations (\ref
{c1}), (\ref{c2}) and (\ref{c3}) defines a K\"{a}hler structure. \ Equation (%
\ref{c1})\ is a compatibility equation between $\Omega _{ab}$ and $g_{ab}$\
, equation (\ref{c2}) is the condition that the metric should be Hermitian,
and equation (\ref{c3}) is the condition that $J_{\ b}^{a}$ should be a
complex structure. \ Let 
\[
\Omega _{ab}=\left( 
\begin{array}{cc}
0 & 1 \\ 
-1 & 0
\end{array}
\right) 
\]
and require that $g_{ab}$\ be a real, symmetric matrix of the form 
\[
g_{ab}=\left( 
\begin{array}{cc}
\hslash g^{(P)} & \quad \cdot \quad \\ 
\cdot & \quad \cdot \quad
\end{array}
\right) . 
\]
Then the solutions\ $g_{ab}$\ and $J_{\ b}^{a}$ to equations (\ref{c1}),(\ref
{c2}) and (\ref{c3}) depend on an arbitrary real function $A$ and are of the
form 
\[
g_{ab}(A)=\left( 
\begin{array}{cc}
\hslash g^{(P)} & A \\ 
A & \qquad \left( \hslash g^{(P)}\right) ^{-1}(1+A^{2})
\end{array}
\right) , 
\]
\[
J_{\ b}^{a}(A)=\left( 
\begin{array}{cc}
A & \qquad \left( \hslash g^{(P)}\right) ^{-1}(1+A^{2}) \\ 
-\hslash g^{(P)} & -A
\end{array}
\right) . 
\]
The choice of $A$ that leads to the simplest K\"{a}hler structure is $A=0$,
which is a unique choice in that it leads to the flat K\"{a}hler metric. \ I
will show this by carrying out the complex transformation that leads to the
canonical form for the flat K\"{a}hler metric. \ I set $A=0$, and work with\
the K\"{a}hler structure given by 
\begin{equation}
\Omega _{ab}=\left( 
\begin{array}{cc}
0 & 1 \\ 
-1 & 0
\end{array}
\right)  \label{SPomega}
\end{equation}
\begin{equation}
g_{ab}=\left( 
\begin{array}{cc}
\hslash g^{(P)} & 0 \\ 
0 & \left( \hslash g^{(P)}\right) ^{-1}
\end{array}
\right)  \label{SPg}
\end{equation}
\begin{equation}
J_{\ b}^{a}=\left( 
\begin{array}{cc}
0 & \left( \hslash g^{(P)}\right) ^{-1} \\ 
-\hslash g^{(P)} & 0
\end{array}
\right)  \label{SPj}
\end{equation}
The complex coordinate transformation is nothing but the Madelung
transformation 
\[
\psi =\sqrt{P}\exp (iS/\hslash ) 
\]
\[
\psi ^{\ast }=\sqrt{P}\exp (-iS/\hslash ) 
\]
In terms of the new variables, (\ref{SPomega}), (\ref{SPg}) and (\ref{SPj})
take the canonical form 
\[
\Omega _{ab}=\left( 
\begin{array}{cc}
0 & i\hslash \\ 
-i\hslash & 0
\end{array}
\right) 
\]
\[
g_{ab}=\left( 
\begin{array}{cc}
0 & \hslash \\ 
\hslash & 0
\end{array}
\right) 
\]
\[
J_{\ b}^{a}=\left( 
\begin{array}{cc}
-i & 0 \\ 
0 & i
\end{array}
\right) 
\]
The Madelung transformation is remarkable in that the Hamiltonian takes the
very simple form 
\[
{\cal H}=\int \left\{ \frac{\hbar ^{2}}{2}g^{\mu \nu }\frac{\partial \psi
^{\ast }}{\partial x^{\mu }}\frac{\partial \psi }{\partial x^{\nu }}+V\psi
^{\ast }\psi \right\} d^{n}x, 
\]
and the equations of motion become linear.

Finally, one introduces a Hilbert space structure using $g_{ab}$,$\ \Omega
_{ab}$\ to define the Dirac product. \ For two wave functions $\phi $, $%
\varphi $ define the Dirac product by 
\begin{eqnarray*}
\left. <\phi |\varphi >\right. &=&\frac{1}{2\hslash }\int \left\{ \left(
\phi (x^{\mu }),\phi ^{\ast }(x^{\mu })\right) \cdot \left[ g+i\Omega \right]
\cdot \left( 
\begin{array}{c}
\varphi (x^{\mu }) \\ 
\varphi ^{\ast }(x^{\mu })
\end{array}
\right) \right\} d^{n}x \\
&=&\frac{1}{2\hslash }\int \left\{ \left( \phi (x^{\mu }),\phi ^{\ast
}(x^{\mu })\right) \left[ \left( 
\begin{array}{cc}
0 & \hslash \\ 
\hslash & 0
\end{array}
\right) +i\left( 
\begin{array}{cc}
0 & i\hslash \\ 
-i\hslash & 0
\end{array}
\right) \right] \left( 
\begin{array}{c}
\varphi (x^{\mu }) \\ 
\varphi ^{\ast }(x^{\mu })
\end{array}
\right) \right\} d^{n}x \\
&=&\int \phi ^{\ast }(x^{\mu })\varphi (x^{\mu })d^{n}x
\end{eqnarray*}
In this way the Hilbert space structure of quantum mechanics results from
combining the symplectic structure of the hydrodynamical model with the
Fisher information metric of information theory.

An important result that comes out of this analysis concerns the issue of
suitable boundry conditions for the fields $P$ and $S$. \ It has been
pointed out \cite{Wallstrom94} that the Schr\"{o}dinger theory is not
strictly equivalent to some of the other formulations (i.e., the
hydrodynamical formulation and stochastic mechanics) because features such
as the quantization of angular momentum, which are natural when the theory
is formulated in terms of wave functions, require an additional constraint
in a theory formulated in terms of hydrodynamical variables. For example, in
the case of the hydrogen atom, the quantization of angular momentum results
from requiring that the wave function be single-valued in configuration
space. \ But the derivation of the K\"{a}hler structure and Hilbert space
structure presented here shows that the Schr\"{o}dinger representation
follows naturally from the hydrodynamical formulation provided we take into
account the role of the Fisher information metric, and furthermore that this
representation is unique in that it is the coordinate system in which the
K\"{a}hler structure takes the simplest form. \ From a purely mathematical
point of view, it is not surprising that the correct boundry conditions are
those that are simplest when formulated in the simplest coordinate system,
i.e. single-valuedness of the canonically conjugate fields $\psi $, $\psi
^{\ast }$. \ 

\section{Appendix}

I want to examine the extremum obtained from the fixed end-point variation
of the Lagrangian $L_{QM}$, equation (\ref{QL}). \ In particular, I wish to
show the following: given $P$\ and $S$\ that satisfy equations (\ref{CONT})\
and (\ref{QHJEQ}), a small variation of the probability density $P(x^{\mu
},t)\rightarrow P(x^{\mu },t)^{\prime }=P(x^{\mu },t)+\epsilon \delta
P(x^{\mu },t)$ for fixed $\sigma $ will lead to an increase in $L_{QM}$, as
well as an increase in the Fisher information $I$.

I assume fixed end-point variations, and variations $\epsilon \delta P$ that
are well defined in the sense that$\ P^{\prime }$ will have the usual
properties required of a probability density (such as $P^{\prime }>0$ and
normalization).

Let $P\rightarrow P^{\prime }=P+\epsilon \delta P$. \ Since $P$\ and $S$\
are solutions of the variational problem, the terms linear in $\epsilon $
vanish. \ If one keeps terms up to order $\epsilon ^{2}$, the change in $%
L_{QM}$ is given by

\begin{eqnarray*}
\Delta L_{QM} &=&L_{QM}(P^{\prime },S)-L_{QM}(P,S) \\
&=&\frac{\epsilon ^{2}\lambda }{2}\int g^{\mu \nu }\left\{ \frac{(\delta
P)^{2}}{P^{3}}\frac{\partial P}{\partial x^{\mu }}\frac{\partial P}{\partial
x^{\upsilon }}-\frac{2(\delta P)}{P^{2}}\frac{\partial P}{\partial x^{\mu }}%
\frac{\partial (\delta P)}{\partial x^{\upsilon }}+\frac{1}{P}\frac{\partial
(\delta P)}{\partial x^{\mu }}\frac{\partial (\delta P)}{\partial
x^{\upsilon }}\right\} dtd^{n}x+O\left( \epsilon ^{3}\right) .
\end{eqnarray*}
Using the relation 
\[
Pg^{\mu \nu }\frac{\partial }{\partial x^{\mu }}\left( \frac{\delta P}{P}%
\right) \frac{\partial }{\partial x\upsilon }\left( \frac{\delta P}{P}%
\right) =g^{\mu \nu }\left\{ \frac{\delta P^{2}}{P^{3}}\frac{\partial P}{%
\partial x^{\mu }}\frac{\partial P}{\partial x^{\upsilon }}-\frac{2\delta P}{%
P^{2}}\frac{\partial P}{\partial x^{\mu }}\frac{\partial \delta P}{\partial
x^{\upsilon }}+\frac{1}{P}\frac{\partial \delta P}{\partial x^{\mu }}\frac{%
\partial \delta P}{\partial x^{\upsilon }}\right\} , 
\]
one can write $\Delta L_{QM}$\ as 
\[
\Delta L_{QM}=\frac{\epsilon ^{2}\lambda }{2}\int P\left\{ g^{\mu \nu }\frac{%
\partial }{\partial x^{\mu }}\left( \frac{\delta P}{P}\right) \frac{\partial 
}{\partial x^{\upsilon }}\left( \frac{\delta P}{P}\right) \right\}
dtd^{n}x+O\left( \epsilon ^{3}\right) , 
\]
which shows that $\Delta L_{QM}>0$ for small variations, and therefore that
the extremum of $\Delta L_{QM}\ $is a minimum. \ Furthermore, since $\Delta
L_{QM}\sim \lambda $, it is the Fisher information term $I$\ in the
Lagrangian $\Delta L_{QM}$ that increases, and the extremum is also a
minimum of the Fisher information.

\end{document}